\documentclass[
 aps,
 pre,
 amsmath,amssymb,
 reprint,superscriptaddress,
showkeys
]{revtex4-2}

\usepackage{epigraph}
\usepackage[T1]{fontenc}
\usepackage[utf8]{inputenc}
\usepackage{amsmath}
\usepackage{amssymb}
\usepackage{graphicx}
\usepackage{refstyle}
\usepackage{skak}
\usepackage{multirow}
\usepackage{epsdice}
\usepackage{soul}
\usepackage[
  citecolor=blue,
  colorlinks,
  linkcolor=blue,
  urlcolor=blue,
]{hyperref}
\usepackage{epstopdf}
\usepackage{bm}
\usepackage[version=4]{mhchem}
\usepackage{tikz}
\usetikzlibrary{arrows,shapes,trees}
\usepackage{booktabs} 
\usepackage{lipsum} 

\begin{document}
	\title{A game played by tandem-running ants: Hint of procedural rationality}
\author{Joy Das Bairagya}
\email{joydas@iitk.ac.in}
\affiliation{
 Department of Physics, Indian Institute of Technology Kanpur, Uttar Pradesh, PIN: 208016, India
}
\author{Udipta Chakraborti}
\email{udi1570@gmail.com}
\affiliation{Present address: Sorbonne Université, CNRS, Inserm, Neuro-SU, 75005, Paris, France
}
\affiliation{Sorbonne Université, CNRS, Inserm, Institut de Biologie Paris-Seine, IBPS, 75005, Paris, France
}
\affiliation{
Behaviour and Ecology Lab, Department of Biological Sciences, Indian Institute of Science Education and Research, Kolkata, Mohanpur 741246, India
}

\author{Sumana Annagiri}
\email{sumana@iiserkol.ac.in}
\affiliation{
Behaviour and Ecology Lab, Department of Biological Sciences, Indian Institute of Science Education and Research, Kolkata, Mohanpur 741246, India
}
\author{Sagar Chakraborty}
\email{sagarc@iitk.ac.in}
\affiliation{
 Department of Physics, Indian Institute of Technology Kanpur, Uttar Pradesh, PIN: 208016, India
}
\begin{abstract} 
Navigation through narrow passages during colony relocation by the tandem-running ants, {\it Diacamma indicum}, is a tour de force of biological traffic coordination. Even on one-lane paths, the ants tactfully manage a bidirectional flow: Informed individuals (termed leaders) guide nest-mates (termed followers) from a suboptimal nest to a new optimal nest, and then return to recruit additional followers. We propose that encounters between the ants moving in opposite directions can be modelled within the framework of game theory leading to an understanding of the mechanism behind observed behaviours. Our experiments reveal that, upon encountering a tandem pair (a leader and its follower) on a narrow path, the returning leader reverses her direction and proceeds toward the new nest again as if she becomes the leader guiding a follower. This observed behaviour is consistent with game-theoretic predictions, provided the assumption of perfect rationality is relaxed in favour of bounded rationality---specifically, procedural rationality. In other words, the experimental outcomes are consistent with sampling equilibrium but not with Nash equilibrium. Our work, which strives to induct the essence of behavioural game theory into the world of ants, is first ever report of realizing sampling equilibrium in scenarios not involving human players.
\end{abstract}

\maketitle
\section{Introduction}
\epigraph{\it Go to the ant, O sluggard; consider her ways, and be wise. Without having any chief, officer, or ruler, she prepares her bread in summer and gathers her food in harvest.}{Proverbs 6:6–8; English Standard Version.}
A longstanding fascination with the life of ants is evident across human societies---as captured in the above biblical exhortation. This admiration, stemming from their self-organized and collective behaviour in the absence of any central command, continues to inspire scientific investigations which sometimes conclude that ants can best even human groups in efficient decision-making~\cite{Dreyer2024}.

Interestingly, the success of classical games in \emph{Homo economicus}~\cite{Persky1995} and behavioural games in \emph{Homo sapiens}~\cite{Spamer1999} in comprehending the epistemological reasons behind the outcomes in  strategic social interactions has not been replicated in the ants. Evidently, ants are cognitively quite capable and their decision-making capabilities---undiminished generations after generations---are obviously genetically hardwired; so, it is quite natural to wonder what ants would do in strategic interactions. The real world natural situations are quite complicated to analyze as there are different kinds of interactions---number of ants involved, number of possible options (actions) available to ants, values of rewards, etc. can vary a lot. In order to develop knowledge about those situations, basic understanding of how ants behave in reduced non-trivial strategic games is required. A survey of literature leaves one surprised that there is a lacunae in the myrmecology and the game theory in this aspect.

We are, thus, motivated to make a novel research inroad in this direction and set up in laboratory a controlled game setup that mimics, in a stylized way, certain natural scenarios that ants would find themselves in their lives. In particular, we focus on an ant species, {\it Diacamma indicum}~\cite{santschi1920,Annagiri2023}, which  do not lay any pheromone trail during nest relocation. (Among the social insects, nest relocation is an essential goal-oriented task, where all colony members need to be moved from an uninhabitable shelter to a new habitable location.) Instead, they employ a recruitment strategy known as tandem running, in which one informed proactive ant---a tandem leader (TL)---guides a single uninformed ant, follower (Fl), towards the new nest while maintaining tactile contact~\cite{sumana2012,Kaur2012}. After resettling the follower into the new nest, she returns to the old nest alone to recruit remaining followers. During her returning journey, the leader is termed as returning leader (RL). We construct a narrow path (having same width as the body width of a {\it D. indicum}) to facilitate bidirectional movements on it---movement of TL {\color{black}and Fl} from the old nest to the new nest and that of RL in the opposite direction---{\color{black}leading to frequent head-on encounters between tandem pairs (TL and Fl) and RLs.} This leads to a \emph{coordination game}~\cite{schelling1980,skyrms2004stag} kind of scenario where TL and RL may want to coordinate their actions to move towards same nest taking Fl with them. Whatever be the theoretical demands of the game, these ants show remarkable adaptive decision-making in the experiments: They quickly mitigate the deadlock by coordinating their movement directions, aligning their subsequent heading directions collectively towards the new nest even in the absence of any enforcing agent (say, policing ants)~\cite{Pathak2023}. In this paper, we wonder about this particular decision-making and reveal a game-theoretic mechanism behind it.

In order to render the novelty of our work conspicuous, we must emphasize that we are \emph{not} using the framework of evolutionary game theory~\cite{SMITH1973,Smithbook1982} which is routinely used to explain certain behaviours in non-human animals, assuming that organisms are myopic in their decision-making and their strategies are genetically inherited rather than a deliberate thought-process. One reason behind this is probably the long-standing belief that only humans can make adaptive decisions, while other animals' behaviours are considered mostly reflexive~\cite {Morgan1984,Hecht2012}. Another reason might just be that it is extremely hard to understand the thought-process of animals with no skill to communicate with human experimentalists. It is, thus, not too surprising that the behavioural game-theoretic studies have remained mostly restricted to humans, chimpanzees and monkeys~\cite{Lee2004,Barraclough2004,Lee2005,Engelmann2017}. 

However, recent studies on social insects challenge this view by demonstrating that non-human animals are capable of adaptive decision-making~\cite{Galef2012,Barber2014}. In particular, ants exhibit remarkable cognitive traits, such as quantitative assessment abilities~\cite{dEttorre2021}, distance measurement~\cite{Wittlinger2007,Pfeffer2016}, associative learning~\cite{Fernandes2017,Chandak2023}, context-dependent behavioural plasticity~\cite{Bey2025}, memory retrieval~\cite{Piqueret2019,Schwarz2011,Huber2018}, individual experience base task switching~\cite{Tanaka2022}, context dependent ant recruitment~\cite{Collignon2009}, and risk sensitivity~\cite{Hbner2017}, pointing to individual-level adaptive decision-making capabilities. Therefore, ants are excellent model organisms for studying behaviour in a game-theoretic context, if at all.

To bring forward another aspect of novelty of our work, we recall that theoretical normative baseline of human behaviour is set by modelling human as \emph{Homo economicus} who are perfectly (von Neumann–Morgenstern) rational~\cite{VNMbook}, strictly self-serving, logico-mathematically intelligent beings always trying to maximize expected utility (their subjective satisfaction; cf.~\cite{page2022optimally}). However, researches in psychology and behavioural economics literature~\cite{kahneman2011thinking,kahneman2021noise,Shafir2002,HOUSER201419,Lee2005} unequivocally establish that real humans (\emph{Homo sapiens}) always differ from rationality~\cite{Touhey1974,ONeill1987,Rapoport1992,Budescu1994,Mookherjee1994,Ochs1995,Binmore2001,Sarin2001,Camerer2010} to act bounded rationally, which assumes unavoidable pragmatic limitations in cognitive abilities and in access to information. A type of bounded rationality is procedural rationality~\cite{Osborne1998} in which an individual's past experiences (or knowledge) about action-consequence relationship in a decision-making scenario fixes the eventually adopted decisions. Recent behavioural studies have found that human behaviours in completely mixed games align more closely with procedural rationality based solutions~\cite{Selten2008}; these observations may have an evolutionary game-theoretic explanation as well~\cite{Bairagya2025}. While it appears quite natural that any animal would be less endowed than a perfectly rational agent, there is no scientific exploration of procedurally rationality in games played by animals. Our work on games in \emph{D. indicum} ants is the first ever reported hint of procedural rationality in non-humans (at least, in invertebrates).

\section{Model: From Experiment to Theory}
\subsection{The observation}
\begin{figure*}
	\centering
	\includegraphics[width=1.0\linewidth]{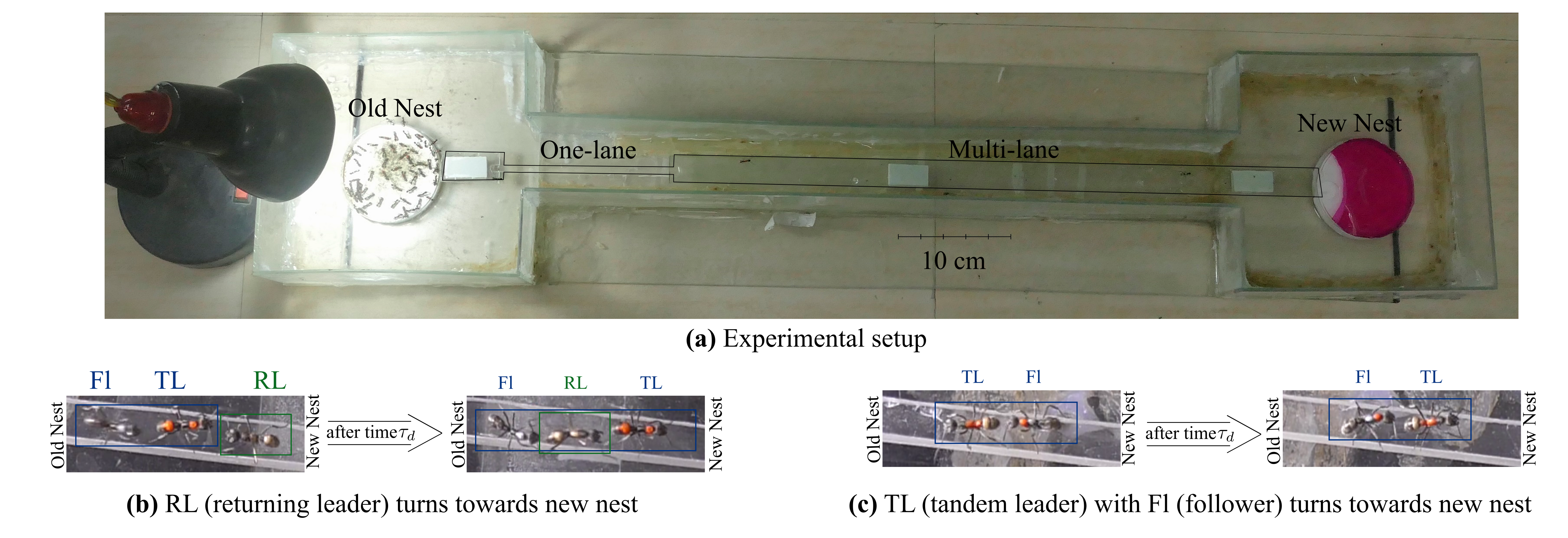}
	\caption{Experimental setup and behavioural observations: Sub-figure (a) depicts the experimental setup showing the relocation path from the old nest to the new nest. A one-lane segment, 14.7 cm long and approximately 0.5 cm wide (approximately the body width of an ant), begins 5.1 cm from the entrance of old nest and is followed by a 62.3 cm multi-lane segment, 5 cm wide. In the alternate arrangement the one-lane segment was placed such that it ended 5.1 cm before the new nest's entrance. The total distance between ON and NN is 77 cm. The entire path was enclosed within a custom-designed water-filled container to prevent ants from escaping. A video camera was focused exclusively on the one-lane region to record behavioural interactions.
	Sub-figure (b) showcases an example of an RL turning back toward the new nest upon encountering a tandem pair. Sub-figure (c) illustrates the ability of turning back by a tandem pair. }
	\label{fig:fig1}
\end{figure*}
Ant colonies were appropriately collected and careful setup was prepared~(see Section~\ref{app:experiments}) to do scientifically meaningful observations. The experiments were carried out at a temperature of 25±2°C, in a rectangular plexiglass arena (100 cm × 40 cm) filled with 2.5 cm of water to prevent ants escaping from the arena (see Fig.~\ref{fig:fig1}a). The artificial nest, made of petri dish with plaster of paris base, containing the ant colony, referred to as the old nest, was placed on one side of the arena, and another nest identical to the old nest was placed opposite side of the arena, termed as new nest. These two nests were connected through a floating plexiglass bridge of 77 cm long---reflecting a typical relocation distance~\cite{Kaur2017}---served as the sole pathway for the relocation event which was initiated by removing the petri dish cover of the old nest and switching on a white light just above ($\sim 15$ cm) the nest. The bridge comprised of two distinct path widths: (i) a multi-lane section (2.5 cm width), allowing more than one ant to walk or move freely side-by-side and (ii) a one-lane section (0.5 cm width), allowing a single ant to walk or move at a time. Majority of the bridge consisted with multi-lane path, however a 14.7 cm long one-lane path was constructed just 5.1 cm before the one end of the bridge. The relocation experiments were performed in two different scenarios: (i) one-lane path near to the new nest, and (ii) one-lane path near to the old nest. To create the two scenarios, we simply reversed the positions of old and new nests.

During the relocation process, a leader takes a nest-mate (or follower) to new nest from old nest. Multiple leaders take part in transporting fellow nest-mates and broods. Since they use the same path, head-on encounters between  TLs and RLs, naturally occurs; if the path is so narrow that most of the times only one individual can pass through it at a time, the decisions taken by the ants at the point of encounter become pivotal for the success of ongoing relocation process. It is very important to realize that both RL and TL are capable of switching their travelling directions if they desire so, as illustrated in Fig.~\ref{fig:fig1}b (see~\href{https://doi.org/10.5281/zenodo.16418136}{video1} in supplemental material) and Fig.~\ref{fig:fig1}c (see~\href{https://doi.org/10.5281/zenodo.16418136}{video2} in supplemental material), respectively. 

So, what strategic decisions are taken by the ants when they are at head-on encounters?  In our experiments, when the followers do not become lost, we observed on the one-lane path, where an RL encounters a tandem pair, either 
\begin{itemize}
\item Observation 1: the RL manages to completely cross the tandem pair side-by-side (see~\href{https://doi.org/10.5281/zenodo.16418136}{video3} in supplemental material), or 
\item \emph{Observation 2: the RL turns back towards the new nest} after trying to force her way towards the old nest for some time (see~\href{https://doi.org/10.5281/zenodo.16418136}{video1} in supplemental material). 
\end{itemize}
The former event is due to experimental setup's limitation in achieving truly one-lane path; the side-by-side crossing is, of course, the norm on multi-lane path.  For the goal of this paper, we want to restrict ourselves to the cases where crossing side-by-side is not present. Thus, we analyzed all the cases where TL and RL encounter on one-lane path---either near the old nest or near the new nest---and documented the encounter position ($l_1$ measured from the old nest) and decision taking time ($\tau_d$) before RL turns. We neither observed a single event where a TL turned towards the old nest to resolve the bottleneck created due to encounter nor we observed resolution of the bottleneck by climbing of one ant over the other. An explanation of aforementioned  \emph{Observation 2} is what we seek and find in this paper.

\subsection{Game theoretical viewpoint}
\label{sec:gtv}
The decision making process at the encounter event between TL and RL is well-suited for adopting game-theoretical concepts of strategic interaction, as outcomes do not depend solely on one leader's decision but on both. Here, we want to emphasize that we do not include any role of Fl in the strategic interaction (although RL and TL physically make tactile contacts with the Fl) as Fl does not make any independent decision regarding her movement direction. This assumption of Fl's relative lack of decision-making is rooted in the observation that upon losing tactile contact, she appears be lost and moves randomly~\cite{Kaur2017, Pathak2023}. Thus, in the most simple yet nontrivial setup of strategic interaction, only the decisions of TL and RL are the determining factors in the outcome of deadlock.

Motivated, we envisage a two-player, one-shot, non-cooperative game. Let us articulate the model mathematical setup of the game more clearly as it helps us to justify the adopted game matrix better. Essentially, we are assuming that there is a lot of leaders in the old nest and the new nest. At any given point of time, there is only one TL and only one RL on the path between the nests. The path is assumed to be a one-dimensional straight line allowing no liberty for the players to pass side by side each other or to be immobile when path in front is clear---forcing the strategy set to be strictly $\{{\rm ON}, {\rm NN}\}$. Here, strategies ON and NN of a player, respectively, refer to the decisions taken by her to head towards the old nest and the new nest. Once both TL and RL reach one of the nests following whatever mutual decision is taken, the game is considered over. Next, the game may be played again between two other leaders in similar manner. The payoff elements are found through contemplation about the question: What are the costs and benefits of the players' choices? The cost can be most simply considered an increasing function of the time spent on the path---more time implies a higher cost. As far as the benefit is concerned, since relocation is directly linked with their fitnesses, the benefit of successful relocation of one follower (sharing players' genes) to the new nest may assumed be to equally shared by both the players. Additionally, when we think more about the situation where mutual decision may not lead to relocation of the follower after one encounter, we realize that in a real colony if their decision solves the path blockage quickly due to possible turning of TL towards old nest, then it yields an indirect benefit by preventing further delays (otherwise, the bottleneck would affect the other leader's transportation-work) and by not letting the follower stuck midway where it may get lost as well. 

In summary, the game is represented in normal form as follows:
\begin{widetext}
	\renewcommand{\arraystretch}{2} 
	\begin{eqnarray}  
		\centering
		\begin{tabular}{cc|c|c|}
			& \multicolumn{1}{c}{} & \multicolumn{2}{c}{{Player 2: RL}}\\
			& \multicolumn{1}{c}{} & \multicolumn{1}{c}{ON} & \multicolumn{1}{c}{\,\,\,\,NN\,\,\,\,}\\\cline{3-4} 
			\multirow{2}*{Player 1: {TL}} 
			& ON 
			& $\epsilon b-\left[\alpha\frac{2l_1}{v_T}+\alpha \tau_d\right],\,\epsilon b-\left[\alpha\frac{l}{v_R}+\alpha \tau_d\right]$ 
			& $\epsilon b-\left[\alpha\frac{2l_1}{v_T}+\alpha \tau_d\right],\,\epsilon b-\left[\alpha\frac{2(l-l_1)}{v_R}+\alpha \tau_d\right]$ \\\cline{3-4}
			
			& NN 
			& $0-\left[\alpha \left(\frac{l_1}{v_T}+\tau\right)+\alpha \tau_d\right],\,0-\left[\alpha\left(\frac{l-l_1}{v_R}+\tau\right)+\alpha \tau_d\right]$ 
			& $b-\left[\alpha\frac{l}{v_T}+\alpha \tau_d\right],\,b-\left[\alpha\frac{(l-l_1)}{v_R}+\alpha\frac{(l-l_1)}{v_T}+\alpha \tau_d\right]$ \\\cline{3-4} 
		\end{tabular}\quad
		\label{eq:payoff-matrix}\nonumber
	\end{eqnarray}
\end{widetext}

where benefit for successfully resettling a follower is $b$ for both TL and RL---as if each of them think that it is she who transported the follower to the new nest. Otherwise, they gain a benefit $\epsilon b<b$  ($\epsilon<1$), say, as a result of mitigating the bottleneck while taking the follower back to the old nest. If they do neither, the benefit is zero. Furthermore, we assume that the cost they incur for moving on the path is linearly proportional to the time spent on the path: Specifically, the cost due to spending time $t$ on the path is $\alpha t$, where $\alpha$ is a positive proportionality constant. If the speed of a TL and an RL---$v_{T}$ and $v_{R}$, respectively; the point of encounter, $l_1$---distance $l_1$ from the old nest; and the total distance from old nest to new nest, $l$, are known, then the amount of time spent in walking starting from a nest to arrival at a nest is easily calculated. With all the above considerations, the entries in the payoff matrix---the ones before commas for player 1 and the ones after for player 2---are conspicuously written in the form of `benefit minus cost'.

\section{Results}
\subsection{Ants are not rational}

To keep the discussion as simple as possible, we assume that a time $\tau_d$ is taken by the players to arrive at their strategic decisions, leading to an additional $\alpha\tau_d$ cost. In our mathematical setup, it suffices to set $\tau_d=0$: The players take instantaneous decision. Finally, of course, if their decisions lead toward a deadlock, they must wait for some additional time, say $\tau$,  (practically very large) theoretically, $\tau\to\infty$. Later when we come back to experiments, we shall see how even relaxing these constraints keeps the theory compatible with the observations.

We now try to find appropriate equilibrium outcome of the strategic game described by the game matrix under consideration. The core question we ask is whether Observation 2 is compatible with any game-theoretic equilibrium concept. To this end, let us start with the classical equilibrium concept, the Nash equilibrium (NE). Suppose any two (von Neumann--Morgenstern) rational players play the game while possessing common knowledge of the game structure and rationality. In that case, each player would opt for a strategy such that she has no incentives to deviate \emph{unilaterally} to adopt alternative strategy; the resultant strategy profile is, technically, called an NE. 

To this end, it turns out to be notationally convenient to denote the payoff elements of TL and RL as $\Pi^{T}_{ij}$ and $\Pi^{R}_{ij}$, respectively, so that the game matrix presented in Sec.~\ref{sec:gtv} can be notationally written in the following short form:
\begin{eqnarray} 
	\renewcommand{\arraystretch}{2} 
	\begin{tabular}{cc|c|c|}
		& \multicolumn{1}{c}{} & \multicolumn{2}{c}{{RL}}\\
		& \multicolumn{1}{c}{} & \multicolumn{1}{c}{ON} & \multicolumn{1}{c}{\,\,\,\,NN\,\,\,\,}\\\cline{3-4} 
		\multirow{2}*{{TL}} & ON & $\Pi^{T}_{11},\,\Pi^{R}_{11}$ & $\Pi^{T}_{12}$,\,$\Pi^{R}_{21}$ \\\cline{3-4}
		& NN & $\Pi^{T}_{21},\,\Pi^{R}_{12} $ & $\Pi^{T}_{22},\,\Pi^{R}_{22}$ \\\cline{3-4} 
	\end{tabular}\quad
	\label{eq:payoff-matrix2}\nonumber
\end{eqnarray}
where the explicit expressions of the elements are obvious. Now, since Observation 2 means that (NN,NN) is the chosen outcome, we look for when (NN,NN) can be NE; essentially, one needs 
\begin{eqnarray}
&&\Pi^{T}_{22}\ge\Pi^{T}_{12} \textrm{~and~} \Pi^{R}_{22}\ge\Pi^{R}_{12},\nonumber\\
\implies&& (1-\epsilon)\frac{b}{\alpha} \ge \frac{l-2l_1}{v_T}~{\rm and}~\frac{b}{\alpha} + \tau \ge \frac{l-l_1}{v_T}\,.
\label{eq:nec}
\end{eqnarray}
Immediate issue is that two parameters, $\frac{b}{\alpha}$ and $\epsilon$, are unknown to us. Nevertheless, we can make a plausible estimate for them as follows.

First, we estimate the minimum value of $b/\alpha$. It is reasonable to assume that the benefit acquired by a solitary ant (say, the first ant who discovered the new nest) is at least equal to the cost of total travel---to and fro between old and new nests---she incurs: In symbols, 
\begin{eqnarray}
b\ge b_{\rm min}\equiv\alpha\frac{l}{v_T}+\alpha\frac{l}{v_R},\label{eq:bmin}
\end{eqnarray}
giving us a lower bound on benefit used in our model.

Next, to get an estimate of $\epsilon$, we realize that a TL has as inherent bias of transporting followers to the new nest. It means that the total payoff of transpiration from the old nest to the new nest while overcoming the encounter must be at least as much as the total payoff achieved over the time a follower is picked from the old nest and taken back to the old nest due to the presence of an obstacle on the path. Mathematically, this means 
\begin{eqnarray}
&&b-\alpha\frac{l}{v_T}\ge\epsilon b-\alpha\frac{2l_1}{v_T}\nonumber\\
\implies&&\epsilon\le\epsilon_{\rm max}\left(\frac{b}{\alpha},l_1\right)\equiv1-\left({\frac{b}{\alpha}}\right)^{-1}\left(\frac{l-2l_1}{v_T}\right). 
\end{eqnarray}
We note 
\begin{eqnarray}
\epsilon_{\rm th}\equiv\min_{\frac{b}{\alpha}}\min_{l_1}\epsilon_{\rm max}\left(\frac{b}{\alpha},l_1\right)=1-\frac{\left(\frac{l}{v_T}\right)}{\left({\frac{b_{\rm min}}{\alpha}}\right)}=\frac{v_T}{v_R+v_T}\qquad\label{eq:eth}
\end{eqnarray} 
is the threshold value of $\epsilon$ such that, irrespective of all allowed values of $b$ and point of encounter ($l_1$), it is advantageous for a TL to head towards the new nest. 

Thus, we have zeroed onto a revised payoff matrix where we fix $b\ge b_{\rm min}$ and $\epsilon\le\epsilon_{\rm th}$. (The ordinal relationship between the payoff elements of the game remains unchanged for game matrices following these two constraints.) Now, we note that expression~(\ref{eq:nec}) is satisfied trivially for encounters at $l_1\ge l/2$ because the r.h.s. of the first inequality is non-positive and the l.h.s. of the second one is very large (as $\tau\to\infty$). Furthermore, for encounters at $l_1< 1/2$, the second inequality is still satisfied trivially and  the first inequality holds for $b\ge b_{\rm min}$ ( Eqs.~\ref{eq:bmin}) and $\epsilon\le\epsilon_{\rm th}$ (Eq.~\ref{eq:eth}). In conclusion, (NN,NN) hold good as NE irrespective of location of encounter. While it may appear that we have just explained Observation 2 assuming the ants are rational, the dilemma is that this theory predicts even (ON,ON) as a possible NE outcome when the encounter is close to the old nest---something we never observed. To see this, we note that the condition that needs to be satisfied is 
\begin{eqnarray}
&&\Pi^{T}_{11}\ge\Pi^{T}_{21} \textrm{~and~} \Pi^{R}_{11}\ge\Pi^{R}_{21},\nonumber\\
\implies&&  \epsilon \frac{b}{\alpha} +\tau  \ge \frac{l_1}{v_T}~{\rm and}~ \frac{l}{2} \ge  l_1  \,,
\label{eq:nec'}
\end{eqnarray}
where the first inequality's l.h.s. is again very large; hence, when encounter point is near to the old nest ($l_1< l/2$), (ON, ON) is a NE.

For most of the readers, the fact that a NE outcome is not witnessed is probably not a surprise---no one considers the ants as rational agents; the epistemic requirements to reach such outcomes is so stringent that even humans seldom opt for decisions that comply with NE. So, it is an intriguing problem to figure out a formalism within the paradigm of game theory such that strategy profile, (NN,NN), turns out to be the theoretically predicted equilibrium; especially, if the formalism allows for theoretical solutions that require less stringent epistemic requirements.

\subsection{Hint of procedural rationality}
In a real setup, either in controlled experiments or in natural habitats, before the nest relocation starts, most of the primary leader ants do extensive exploration~\cite{Mukhopadhyay2019} for new nest---this is called discovery phase in the relocation process. These ants assess the path from old nest to new nest (even multiple new nests if were available) and gather knowledge about the nests' suitability~\cite{Karunakaran2017}, direction and distance. During the discovery (and even later during relocation), these ants invariably interact with other ants travelling in the opposite direction. It goes without saying that the ants dynamically gather information using direct, although maybe incomplete, experience. Since the payoffs (as described in the payoff matrix above) are effected only by the travel times and transport of a follower ant, it is fair to assume that ants have a fair idea of the payoffs to be realized. Hence, it is tempting to hypothesize that for the problem in hand, an ant may act \emph{as if} it is a procedurally rational player who learns through \emph{best experienced payoff dynamics}~\cite{Sandholm2020}---i.e., a learning dynamics where strategy is chosen based on highest payoff experienced while sampling all the strategies. Specifically, we consider players who independently sample the strategies available to them \emph{once} and pick the one which leads to better consequence in terms of payoff (any tie is broken equiprobably)---such a process, known as payoff \emph{sampling dynamics}, would allow for \emph{sampling equilibrium} strategy profile as one of its fixed points. Naturally, the stable ones are expected to be realized.

In one interpretation of the aforementioned sampling process, it is assumed that the procedurally rational player indulges in a
mental process to build action-consequence associations. The mental process may seek inputs from experiences in similar past games and observations of other players. Also, if the ancestors of players evolved in an environment which involved similar games, then natural selection could lead to the descendants behaving as if they know information about the game; in other words, a player's behaviour is adapted to its ancestors' environment. Basically, in certain situations, like in ants---where generation after generation relocation processes have been happening---the mental process may even be considered as genetically hardwired.

Once we adopt the above view, rest of the arguments follows straightforwardly. Let us consider two procedurally rational players playing the above-described game and suppose at their respective $n$-th encounter, the TL and the RL choose strategy ON with probability, $x_n$ and $y_n$, respectively. First, we consider the cases where $0<l_1 < l/2$. The ordinal relationship between the payoff elements of TL is
\begin{eqnarray}
\Pi^{T}_{22}>\Pi^{T}_{11}=\Pi^{T}_{12}>\Pi^{T}_{21}.\label{eq:payofftl}
\end{eqnarray} 
Similarly, the ordinal relationship between the payoff elements of RL is
\begin{eqnarray}
	\Pi^{R}_{22}>\Pi^{R}_{11}>\Pi^{R}_{21}>\Pi^{R}_{12}.\label{eq:payoffrl}
\end{eqnarray}
When TL samples ON, she experiences payoff $\Pi^{T}_{11}$ with probability
$y_n$ and payoff $\Pi^{T}_{12}$ with probability $1-y_n$. Likewise, when TL samples NN, she experiences payoff $\Pi^{T}_{21}$ with probability
$y_n$ and payoff $\Pi^{T}_{22}$ with probability $1-y_n$. After this sampling (possibly as a mental process), she has to infer which what probability (i.e., $x_{n+1}$) she should play ON strategy during $(n+1)$-th encounter. Obviously, this probability is equal to the probability with TL finds ON fetching higher payoff after the above-mentioned independent once sampling of each strategy. This probability naturally depends on the probability with which RL mixes her strategies. Therefore, given relation (\ref{eq:payofftl}) above, the probability that the consequence TL associates with ON is superior to the consequence she associates with NN is $y_n\cdot y_n$ (when sampling ON leads to $\Pi^{T}_{11}$ and sampling NN leads to $\Pi^{T}_{21}$) plus $(1-y_n)\cdot y_n$ (when sampling ON leads to $\Pi^{T}_{12}$ and sampling NN leads to $\Pi^{T}_{21}$). Consequently, 
	\begin{eqnarray}
		x_{n+1} = y^2_{n} + y_{n}(1-y_{n}) = y_{n}.\label{eq:sdx}
	\end{eqnarray}  
Similar considerations for RL leads to
	\begin{eqnarray}
	y_{n+1} = x^2_{n}.\label{eq:sdy}
\end{eqnarray} 
Equations (\ref{eq:sdx}) and (\ref{eq:sdy}) form the sampling dynamic for the process in hand; it has two fixed points---$(x^*,y^*)=(0,0)$ and $(1,1)$---which are sampling equilibria as well. However, only $(0,0)$, which corresponds to both RL's and TL's heading towards the new nest, is \emph{stable sampling equilibrium}. 

The cases where $l>l_1 > l/2$ can be analyzed likewise. The ordinal relationship between the payoff elements of TL remains as in relation~(\ref{eq:payofftl}), but the ordinal relationship between the payoff elements of RL changes to the form of relation~(\ref{eq:payofftl}) with locations of $\Pi^{R}_{11}$ and $\Pi^{R}_{21}$ swapped. The sampling dynamic turns out to be
	\begin{eqnarray}
	x_{n+1} = y_{n};\quad y_{n+1} = 0,\label{eq:sdxy2}
\end{eqnarray} 
which yields $(x^*,y^*)=(0,0)$ as the sole stable sampling equilibrium.

In conclusion, if one aptly uses a procedurally rational agent to model ants, then Observation 2 is easily explained as the unique stable outcome of the corresponding game dynamics.
\subsection{From theory back to experiment}

Having gained the game-theoretic insight into the exclusive turning of RLs towards the new nest irrespective of the location of encounter, we would like to break somewhat free from the idealized payoff matrix used in the theoretical analysis. Specifically, we would like to come back to the rather strict assumptions---$\tau_d=0$ and $\tau\to\infty$---as the decision-taking time is bound to be non-zero and no ant would remain stuck in an encounter till her death. Moreover, while trying to go beyond these convenient mathematical constraints, we also take the opportunity to check compatibility between our theory and experiment.

To this end, we realize that $\tau_d$ can be observed directed in experiments---at least when strategy profile (NN,NN) is in action because, once we have eliminated all the side-by-side passing of the ants from our consideration, only Observation 2 has been witnessed. Of course, in experiments, $\tau_d$ is a random variable and so is the point of encounter, whose distance from the old nest is $l_1$. In our preceding mathematical treatment, very high values of $\tau$ may be seen to associate too much cost to the TL's adopting ON strategy. Therefore, we want to understand if a finite low value of $\tau$ is enough to render the theoretical results match with Observation 2. In principle, $\tau_d$ and $\tau$ (hence, $\lambda$) are two random variables (possibly independent); but unlike $\tau_d$, we do not have any experimental estimation of $\tau$. Thus, let us express $\tau$ in the units of $\tau_d$ at each encounter, i.e., $\tau = \lambda \tau_d$, where $\lambda$ is assumes to be a constant non-negative number as a simplifying assumption. This assumption is sufficient to illustrate that even a very tiny fraction of $\tau_d$ as $\tau$ is enough for Observation 2.

\begin{figure}[h!]
	\centering
	\includegraphics[width=1.0\linewidth]{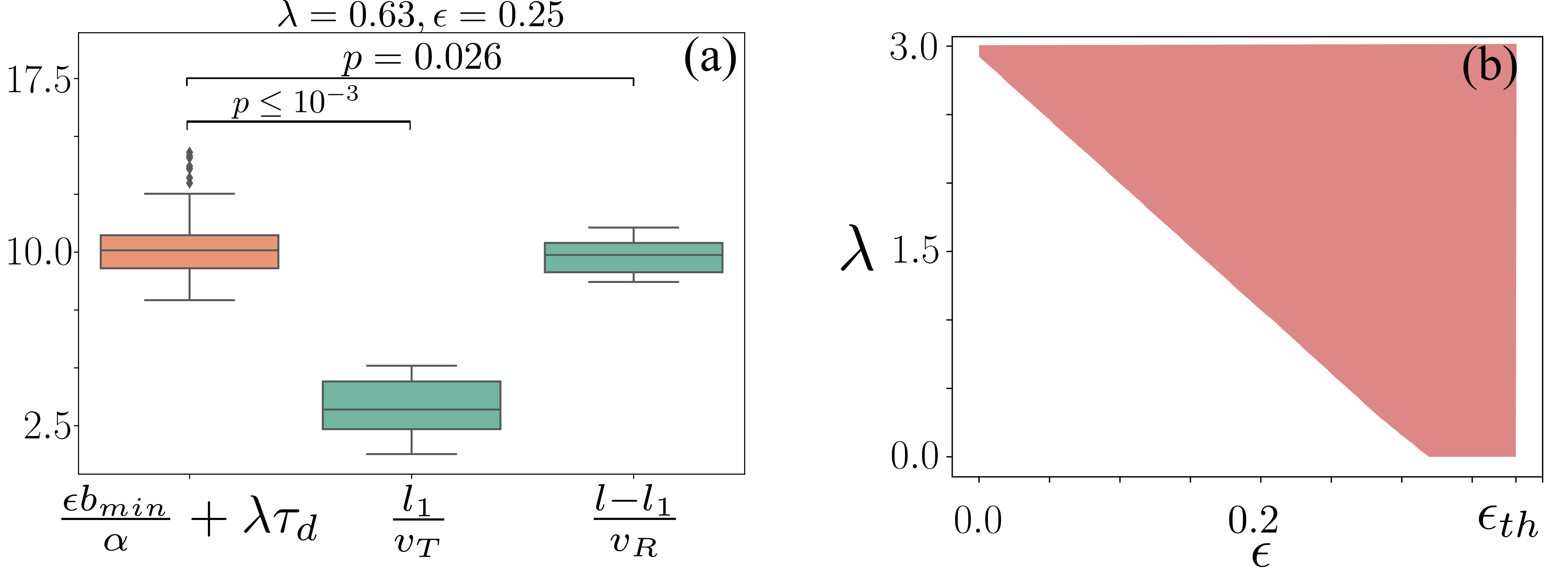}
	\caption{\color{black} Statistical verification of sampling equilibrium conditions. Subfigure (a), is a box-whisker plot, illustrates though a test case that there exists a value of $\lambda$ and $\epsilon$ for which the inequalities, $\epsilon \frac{b}{\alpha} + \lambda \tau_d > \frac{l_1}{v_T}$ and $\epsilon \frac{b}{\alpha} + \lambda \tau_d > \frac{l - l_1}{v_R}$, required for (NN,NN) to be stable sampling equilibrium irrespective of the encounter point, are statistically satisfied, thereby, supporting Observation~2 as a sampling equilibrium. Here, $b$ has been fixed to $b_{\rm min}$. We have used the Wilcoxon signed-rank (significance level = 0.05) to ascertain whether the statistical distributions of the both sides of the above inequalities differ significantly. As indicated in the figure, both tests yielded $p$-values below the significance threshold, leading to rejection of the null hypothesis and supporting the validity of the inequalities.  Subfigure (b) showcases the range (red region) of $\lambda$-values for which the inequalities hold, across different values of $\epsilon\le\epsilon_{\rm th}$.}
	\label{fig:2}
\end{figure}

Our theoretical model assisted by statistical inferencing (see Fig.~\ref{fig:2} and Sec.~\ref{sec:tec} for details) suggests, there exists a set of $\epsilon \le \epsilon_{\rm th}$ for which the minimum waiting time ($\tau_{\rm min}=\lambda_{\rm min}\tau_d$) required for (NN,NN) to become a stable sampling equilibrium is zero, i.e., $\lambda_{\rm min}=0$. This result implies that for a sufficiently high benefit of mitigating the bottleneck, the returning leader---as a procedural rational agent---will always choose to turn towards the new nest.
	
%
%
\section{Discussion } 
We have discovered that the deadlocks resulting during the encounters between two \emph{D. indicum} ants moving in opposite directions on a narrow path during nest relocation can be effectively modelled as a two-player, two-strategy coordination game. Our experiments reveal that the ants’ behaviour aligns with the equilibrium concept of stable sampling equilibrium suggesting that their cognitive limitation is a very important factor. The ants behave as if they are procedurally rational agents. We reiterate that application of (non-evolutionary) game theory in an invertebrate and finding hint of sampling equilibrium in non-humans are the original contributions of this paper which, therefore, advocates for behavioural game theory as a versatile tool for explaining behaviours across species, including both humans and non-human animals. 
	
Our findings once again~\cite{Holldobler1994} challenge the long-standing notion of ant societies as mechanistic systems, where individuals are genetically programmed to perform the same task repeatedly---what has been described as a “factory constructed inside a fortress”~\cite{Holldobler1994}. Instead, we demonstrate that ants are akin to agents capable of adaptive decision-making: They can abandon a previously assigned role and choose strategies appropriate to the immediate context. Specifically, we observed that an RL, en route the old nest, turns back toward the new nest---rather than forcing her way to the old nest---upon encountering a tandem pair in a narrow path, thereby, effectively resolving the deadlock caused by the spatial constraint.

There is no denying the fact that this study has been done by keeping the mathematical model non-trivial and yet as simple as possible while capturing the most important essence of the experimental observations. Therefore, although our study offers novel insights into the applicability of game theory to ant behaviour, some challenges and limitations that need to be addressed in future must be acknowledged.
 
First, the narrow paths used in the experiment are not strictly one-lane. As a result, in many encounters, ants were able to completely pass each other side-by-side---a behaviour we have referred to as \emph{Observation 1}. This outcome has not been captured by our current game-theoretic model, which assumes a strictly one-lane path and, thus, excludes such actions. If one would like to capture the full range of observed behaviours, one would need to include side-by-side crossing as an additional strategy, resulting in a more complex three-strategy, two-player game. However, such models are analytically more difficult and require more experiments to gain additional insights.  
 
Second, like all behavioural studies, we cannot directly observe the cognitive processes underlying the ants’ decisions. Our adoption of procedural rationality is based on the assumption that ants use prior experience to choose their relocation direction, which we modelled using best-experience dynamics. While this framework explains the observed outcomes, it must only be an approximation of the actual decision-making mechanism. Future studies focusing on the exploration phase---particularly the period in which ants thoroughly inspect multiple nest options prior to initiating nest relocation---will offer deeper comprehension of their cognitive strategies at the behavioural level. Furthermore, we have restricted attention to the case where each procedurally rational agent samples every action only once independently. We have not delved into the scenarios where sampling is done $k$ times ($k>1$) for each action. Such an exercise would merely make the presentation more opaque because analytical expression would be far cumbersome. However, one fact is certain, given the fact that $k\to\infty$ forces sampling equilibrium to converge to NE~{\cite{Osborne1998}} (which cannot explain the observed behaviour): There should be a thresold value of $k$ up to which sampling equilibrium should be compatible with the observed behaviour. More interesting scenario involves the case when the actions are not sampled symmetrically---one action is sampled more times or with more probability than the other---something living beings would often do. Intriguingly, such a framework remains theoretically unstudied even in case of human players, to the best of our knowledge. 

Third, a curious reader could claim that the decision making during encounters on the one-lane path actually involved three ants---let us not forget the follower (Fl) ant. So could it be that what we reported in this paper is better explained by treating the outcome as a collective decision-making under a cooperative game~\cite{VNMbook}, unlike the non-cooperative, one-shot, bimatrix game framework adopted herein. Or, at least, maybe one could treat the scenario as a sequential extensive game~\cite{Kuhn1950,bonanno2018game} involving two steps: first RL plays with TL and then she plays with follower. Research projects along these directions is not just based on hypothetical but on the observation that when RL meets TL, and they decide to adopt (NN,NN) as the strategy profile, the trio that travels towards the new nest mostly has the sequence: TL--RL--Fl (with TL leading) and not RL--TL--Fl (with RL leading), naturally suggesting that a game between RL and Fl is on the cards.
 
 It will be of interest to extend the two-ant interaction games to more elaborately model entire multi-stage process through which rest relocation is effected by tandem-runs without any centralized command. Such an endeavour may lead to incorporation of game-theoretic ideas into stochastic drift-diffusion models~\cite{DC2021}. We conclude by highlighting that our study focused on tandem-running ants ({\it D. indicum}), that do not rely on pheromone trails during relocation. Whether pheromone-laying ants can also be modelled effectively within a game-theoretic framework---perhaps under a cooperative game paradigm where decisions emerge at the colony level---remains an open question. Exploring such cases may expand the applicability of behavioural game theory to broader forms of collective animal behaviour.
\section{Materials and Methods}
\subsection{Experimental Protocols}
\label{app:experiments}
A total of sixteen colonies of \emph{Diacamma indicum} (Santchi, 1920) were collected using the nest flooding method~\cite{Kaur2012} from the IISER Kolkata campus at Mohanpur, Nadia, West Bengal, India (22°56´N, 88°31´E) during November 2024 to February 2025. All the colonies were maintained in laboratory conditions following a standard methodology~\cite{Mukhopadhyay2019}, along with the ad libitum food (ant cake) and water~\cite{Bhatkar1970}. Only colonies comprising a gamergate (the only potential reproductive female) were used for the experiment; thus, after the colony collection, each of them was examined using a stereomicroscope to search for gamergate. Alongside this, additional information such as the total number of female individuals, total numbers of broods (egg, larva, pupae), and numbers of males were also recorded. For providing unique identification, all the colony members were marked with a unique colour code by using non-toxic enamel paint (Testors, Rockford, IL, USA).

{\color{black}Six colonies having 99.0±10.9 (mean$\pm$SD) adult females, used in the set of olpNN (one-lane path near to the new nest ) and another six colonies consisting of 102.3±9.9 adult females, used in the olpON (one-lane path near to the old nest) set. The colony size of olpON and olpNN sets were comparable (Mann--Whitney U test=24, p value=0.56).}
The experiment was started by placing the old nest in the arena and remain undistrubed for 15 minutes, to acclamitze them in the arena environment. After 15 minutes, the relocation event was initiated by removing the nest roof (here petri dish cover) of the old nest and providing a white light ~15 cm above the nest to motivate the colony for relocation. Following the last leader follow leader (LFL) tandem run~\cite{Kaur2012} and the transfer of all the brood item, we waited for 15 minutes to ensure any additional transportation and then after considered as the end of relocation process. After each relocation event, the path of the bridge as well as the arena walls was thoroughly cleaned by using ethanol to avoid the influence of any chemical substance that may left behind by the colony. We provided a gap at least of 24 hours between two relocation events. All the experiments were recorded using a video camera  precisely focusing on the one-lane path of the bridge (Panasonic, model: HC-V270) in 25 fps.
\subsection{Theory-Experiment Compatibility}
\label{sec:tec}
Since our theoretical results remain intact for all positive $\epsilon\le\epsilon_{\rm th}$ and $b\ge b_{\rm min}$; hence, in what follows we restrict ourselves to such values of $\epsilon$ and $b=b_{\rm min}$. 

When $l_1<\frac{l}{2}$, among the inequalities given in relations~(\ref{eq:payofftl})~and~(\ref{eq:payoffrl}), following conditions must be checked for their validity for (NN,NN) to be stable sampling equilibrium: $\Pi^{T}_{12} > \Pi^{T}_{21}$ and $\Pi^{R}_{12} < \Pi^{R}_{21}$; other implies inequalities relations~(\ref{eq:payofftl})~and~(\ref{eq:payoffrl}) are either always satisfied or they are not required for finding stable SE (e.g., comparison between $\Pi^{T}_{11}$ and $\Pi^{T}_{12}$, etc.). The above two conditions translate, respectively, to
\begin{subequations}
\label{eq:cond}
\begin{eqnarray}
	&&\epsilon \frac{b_{\rm min}}{\alpha} + \lambda \tau_d > \frac{l_1}{v_T},\\
	&&\epsilon \frac{b_{\rm min}}{\alpha} + \lambda \tau_d > \frac{l - l_1}{v_R}.
\end{eqnarray}
\end{subequations}
Now, we need to find the range of $\lambda$, if at all one exists, that makes two inequalities hold true in experiments.

For this purpose, we conduct a  Wilcoxon signed-rank test between the R.H.S. and the L.H.S. of two inequalities~(\ref{eq:cond}), choosing a significance level of 0.05. As an illustration, see Fig.~\ref{fig:2}a: We collect all the incidences of encounters at any $l_1<\frac{l}{2}$, and record respective values of $l_1$ and $\tau_d$. These events are presented in the box-whisker plots. Next we try to find whether there exists the pair of free parameters, $\epsilon$ and $\lambda$, such that the inequalities holds true. For illustrative purpose, we fix $\epsilon = 0.25$ (which is less than $\epsilon_{\rm th}\equiv\frac{v_T}{v_R + v_T}=\frac{4.0}{6.5 + 4.0} \approx 0.381$, on using reported typical speeds \cite{Kaur2017} of TL and RL) and do trial-and-error to find a $\lambda$, so that two inequalities~(\ref{eq:cond}) can be said to have been satisfied statistically. We find $\lambda=0.63$ is one such value. The existence of $\lambda$ means that even with a finite waiting time, the theory ( which mathematically used infinite waiting time) is capable of explaining Observation 2. In Fig.~\ref{fig:2}(b), we present all possible valid pairs of $\epsilon$ and $\lambda$ (see green coloured region).

As far as the encounters at $l_1 > \frac{l}{2}$ are concerned, the following inequalities always hold: $\Pi^T_{11} = \Pi^T_{12} < \Pi^T_{22}$, $\Pi^R_{11} < \Pi^R_{22}$, $\Pi^R_{11} < \Pi^R_{21}$, and $\Pi^R_{12} < \Pi^R_{22}$. Thus, apart from the dynamics described by Eq.~(\ref{eq:sdxy2}) (which strictly hold when $\tau_d=0$ and $\tau\to\infty$), as listed below, the other three types of dynamics are possible (having relaxed the requirements---$\tau_d=0$ and $\tau\to\infty$) when $\tau_d$ is non-zero and finite, and $\lambda$ is finite:
	\begin{enumerate}
		\item If the ordinal relationship between $\Pi^T_{11} (= \Pi^T_{12})$ and $\Pi^T_{21}$ changes to $\Pi^T_{11} = \Pi^T_{12} < \Pi^T_{21}$, while other relationships specified in Eq.~(\ref{eq:payoffrl}) for RL remain unchanged, then Eq.~(\ref{eq:sdxy2}) modifies to:
		\begin{eqnarray}
			x_{n+1} = 0; \quad y_{n+1} = 0.\label{eq:sdxy2_1}
		\end{eqnarray}
		\item If the ordinal relationship between $\Pi^T_{11} (= \Pi^T_{12})$ and $\Pi^T_{21}$ remains unchanged but the relationship between $\Pi^R_{12}$ and $\Pi^R_{21}$ in Eq.~(\ref{eq:payoffrl}) changes to $\Pi^R_{21}<\Pi^R_{12}$, then the dynamics are given by:
		\begin{eqnarray}
			x_{n+1} = y_n; \quad y_{n+1} = (1 - x_n)x_n. \label{eq:sdxy2_2}
		\end{eqnarray}
		\item If the ordinal relationship between $\Pi^T_{11}(= \Pi^T_{12})$ and $\Pi^T_{21}$ changes to $\Pi^T_{11} = \Pi^T_{12} < \Pi^T_{21}$ and, in addition, the relationship between $\Pi^R_{12}$ and $\Pi^R_{21}$ changes to $\Pi^R_{21}<\Pi^R_{12}$, then the dynamic becomes:
		\begin{eqnarray}
			x_{n+1} = 0; \quad y_{n+1} = (1 - x_n)x_n. \label{eq:sdxy2_3}
		\end{eqnarray}
	\end{enumerate}
In all three cases described by Eqs.~(\ref{eq:sdxy2_1}),~(\ref{eq:sdxy2_2}), and~(\ref{eq:sdxy2_3}), $(0,0)$ happens to be the only stable sampling equilibrium. Therefore, our theoretical conclusion is consistently aligned with Observation~2 when $l_1 > \frac{l}{2}$ for any $\lambda$ given any non-zero $\tau_d$ as would be seen in a real scenario. The compatibility with experiments have been exhibited in Fig.~\ref{fig:2}.
\section*{Data availability statement}
All data analyzed in this paper are available at Zenodo:~\href{https://doi.org/10.5281/zenodo.16417516}{https://doi.org/10.5281/zenodo.16417516}.\\
All codes used to generate the plot in this paper are available at Github:~\href{https://github.com/joydasbairagya/A-game-played-by-tandem-running-ants-Hint-of-procedural-rationality}{https://github.com/joydasbairagya/A-game-played-by-tandem-running-ants-Hint-of-procedural-rationality}. \\
All videos mentioned in the paper are available at Zenodo:~\href{https://doi.org/10.5281/zenodo.16418136}{https://doi.org/10.5281/zenodo.16418136}
\section*{Conflicts of interest statement}
The authors report there are no competing interests to declare.
\acknowledgements
The authors thank Srinivas Arigapudi for insightful comments on the manuscript. JDB has been supported by Prime Minister's Research fellowship (govt. of India).
\bibliography{Bairagya_ant}
\end{document}